\documentclass[journal=jctcce,manuscript=article]{achemso}
\usepackage{amsmath}
\usepackage{amssymb}
\usepackage[version=4]{mhchem}
\usepackage{xcolor}
\usepackage{hyperref}
\newcommand{\ket}[1]{ \left| #1 \right \rangle}

\newcommand{\comm}[2]{ \left[ #1 , #2 \right ]}

\author{Peter Reinholdt}
\affiliation[SDU]{Department of Physics, Chemistry and Pharmacy, University of Southern Denmark, Campusvej~55, DK--5230 Odense M, Denmark}
\email{reinholdt@sdu.dk}
\author{Erik Rosendahl Kjellgren}
\affiliation[SDU]{Department of Physics, Chemistry and Pharmacy, University of Southern Denmark, Campusvej~55, DK--5230 Odense M, Denmark}
\author{Juliane Holst Fuglsbjerg}
\affiliation[KU]{Department of Chemistry, University of Copenhagen, DK-2100 Copenhagen \O.}
\author{Karl Michael Ziems}
\affiliation[DTU]{Department of Chemistry, Technical University of Denmark, Kemitorvet Building 207, DK-2800 Kongens Lyngby, Denmark.}
\author{Sonia Coriani}
\affiliation[DTU]{Department of Chemistry, Technical University of Denmark, Kemitorvet Building 207, DK-2800 Kongens Lyngby, Denmark.}
\author{Stephan P. A. Sauer}
\affiliation[KU]{Department of Chemistry, University of Copenhagen, DK-2100 Copenhagen \O.}
\author{Jacob Kongsted}
\affiliation[SDU]{Department of Physics, Chemistry and Pharmacy, University of Southern Denmark, Campusvej~55, DK--5230 Odense M, Denmark}

\title{Subspace methods for the simulation of molecular response properties on a quantum computer}

\begin{document}

\begin{tocentry}
\end{tocentry}

\begin{abstract}
We explore Davidson methods for obtaining excitation energies and other linear response properties within quantum self-consistent linear response (q-sc-LR) theory.
Davidson-type methods allow for obtaining only a few selected excitation energies without explicitly constructing the electronic Hessian since they only require the ability to perform Hessian-vector multiplications.
We apply the Davidson method to calculate the excitation energies of hydrogen chains (up to \ce{H10}) and analyze aspects of statistical noise for computing excitation energies on quantum simulators.
Additionally, we apply Davidson methods for computing linear response properties such as static polarizabilities for \ce{H2}, \ce{LiH}, \ce{H2O}, \ce{OH-}, and \ce{NH3}, and show that unitary coupled cluster outperforms classical projected coupled cluster for molecular systems with strong correlation.
Finally, we formulate the Davidson method for damped (complex) linear response, with application to the nitrogen K-edge X-ray absorption of ammonia, and the $C_6$ coefficients of \ce{H2}, \ce{LiH}, \ce{H2O}, \ce{OH-}, and \ce{NH3}.
\end{abstract}

\section{Introduction}

Quantum computing is an emerging technology that may solve certain computational problems significantly faster than classical computer architectures.
One of the most promising areas of application is within quantum chemistry, where quantum computers have been proposed for solving the electronic Schr{\"o}dinger equation.~\cite{bauer2020qainqc,cao2019qcinqc,motta2022emerging} 
However, current available quantum hardware, i.e., near-term noisy intermediate-scale quantum (NISQ) devices, are fairly limited in their usefulness due to relatively small qubit counts and their susceptibility to noise, which prevents meaningful execution of deep quantum circuits required for the most promising quantum algorithms.~\cite{nash2020quantum,lau2022nisq}

The variational quantum eigensolver (VQE) approach\cite{peruzzo2014variational,mcclean2016theory} attempts to circumvent this problem with a hybrid quantum-classical workflow in which quantum and classical processors cooperate. In this approach, the quantum processor evaluates relatively shallow quantum circuits, mainly involving preparing quantum states and evaluating expectation values over certain quantum mechanical operators (most notably the Hamiltonian) while the classical computer is used for obtaining the fermionic Hamiltonian and updating the parameters of the prepared wave function. The VQE method estimates the energy from the Rayleigh quotient as
\begin{equation}
    E(\theta) = \frac{\left<\Psi(\theta)\left|\hat{H}\right|\Psi(\theta)\right>} {\left<\Psi(\theta)\left.\right|\Psi(\theta)\right>}
    \label{eq:rayleigh}
\end{equation}
where the wave function depends on a set of parameters (or angles) $\theta$. 
The quantum processor is used to prepare a state $\ket{\Psi(\theta)}$ and to measure terms of the Hamiltonian. The classical processor is then used to collect and post-process the obtained results and to update the parameters $\theta$.

Much attention has been paid to obtaining ground-state energies with the VQE approach. However, many molecular properties require a description of the excited states. For example, absorption spectra can be obtained from excitation energies and oscillator strengths of a molecular system.
Several approaches for obtaining excitation energies have emerged in recent years\cite{motta2020determining,sun2021quantum,mcclean2017hybrid,ollitrault2020quantum,nakanishi2019subspace,higgott2019variational,ziems2023options,jensen2024quantum}.
In this work, we will focus on the quantum linear response (qLR) method, which is similar to the quantum equation of motion (qEOM) of \citeauthor{ollitrault2020quantum}\cite{ollitrault2020quantum}. The qEOM method is an extension of the well-known classical EOM\cite{rowe_equations-of-motion_1968,schaefer_equations_1977,bartlett2012coupled} approach that can potentially leverage near-term, noisy quantum hardware to obtain excitation energies of molecules.
After preparing and optimizing a ground state using VQE, the method uses a quantum processor to measure individual matrix elements of an EOM secular equation. The LR equations are then solved on a classical computer.
The qEOM method was refined by \citeauthor{asthana2023quantum} with the q-sc-EOM\cite{asthana2023quantum} method, which introduced a self-consistent excitation operator manifold originally developed by \citeauthor{prasad1985sc}\cite{prasad1985sc} in the context
of self-consistent propagator theories. These operators satisfy the vacuum annihilation condition and were shown to provide more accurate excitation energies, ionization potentials, and electron affinities than the original qEOM approach\cite{asthana2023quantum}.
The self-consistent operators have also been applied to obtain linear response properties within the self-consistent quantum linear response q-sc-LR approach \cite{kumar2023quantum}. Recently, some of us have also explored the feasibility of other operator transformations for qLR formalisms\cite{ziems2023options}.

When targeting larger molecules, the explicit construction of the Hessian can become a computational bottleneck in terms of quantum resources since many matrix elements need to be measured.  Eventually, the classical processing costs may also become significant when the cost of matrix diagonalization becomes prohibitive.
The same problem is faced in many classical linear response approaches, where Davidson-type methods\cite{DAVIDSON197587} are commonly applied to overcome this issue.
Resorting to Davidson-type approaches allows us to obtain excitation energies (and linear response properties) without explicitly constructing the electronic Hessian, as they only require the ability to perform Hessian-vector multiplications.

The Davidson method\cite{DAVIDSON197587} has already been explored with q-sc-EOM by \citeauthor{kim2023two}\cite{kim2023two}, where it was used for the simulation of excitation energies of \ce{H2}, \ce{H4}, \ce{LiH}, and (frozen-core) \ce{H2O} molecules in an STO-3G basis.
In the work by \citeauthor{kim2023two}, elements of the subspace matrix are directly measured (without the addition of ancillary qubits), which entails the evaluation of transition matrix elements of the UCC Hamiltonian between guess vectors and excited Slater determinants. 
The implementation in Ref.~\citenum{kim2023two} operates and manipulates state vectors on a classical computer. At first glance, this 
might appear problematic since the dimension of state vectors scales exponentially with the number of qubits. However, by using sparse vectors and truncating the excitation level in the EOM expansion, the number of classically stored elements is only polynomial. 

In this work, we use an approach that does not require explicitly computing transition matrix elements of the UCC Hamiltonian.
Instead, we formulate the Davidson method in terms of Hessian-vector products, which only require gradients over excitation operators.
We investigate using Davidson methods for obtaining excitation energies and implement a Davidson solver for linear response properties within the quantum self-consistent linear response unitary coupled cluster approach.

The rest of the paper is organized as follows. In section \ref{sec:theory}, we briefly review the fundamental theory describing the VQE method for unitary coupled cluster. Next, we outline the central equations in q-sc-LR and formulate the associated Davidson method in terms of Hessian-vector multiplications for both excitation energies in section \ref{sec:theory_excitation_energy}, linear response properties in section \ref{sec:theory_lr}, and damped response theory in \ref{sec:theory_damped}.
In section \ref{sec:computational_details}, we provide computational details. 
In section \ref{sec:results_excitation_energy}, we report on the excitation energies of hydrogen chains (up to \ce{H10}) and analyze aspects of the statistical noise in section \ref{sec:results_noise}.
In section \ref{sec:results_polarizability}, we compute static polarizabilities for \ce{H2}, \ce{LiH}, \ce{H2O}, \ce{OH-}, and \ce{NH3}, and show that unitary coupled cluster outperforms classical projected coupled cluster for molecular systems with strong static correlation.
Finally, in section \ref{sec:results_dynamic_polarizability}, we use our formulation of the Davidson method for damped (complex) linear response, with application on the nitrogen K-edge X-ray absorption of ammonia, while we compute $C_6$ coefficients in section \ref{sec:results_c6} (which entail computations of the real electric dipole polarizability at imaginary frequencies) of \ce{H2}, \ce{LiH}, \ce{H2O}, \ce{OH-}, and \ce{NH3}.

\section{Theory}\label{sec:theory}

The unitary coupled cluster wave function is parametrized as
\begin{equation}
    \ket{\Psi_\textrm{UCC}} = e^{\hat{T}(\boldsymbol{\theta})-\hat{T}^{\dagger}(\boldsymbol{\theta})}\ket{0}
\end{equation}
with the cluster operator, $\hat{T}$, being defined as:
\begin{equation}
    \hat{T}(\boldsymbol{\theta}) = \sum_n\ \hat{T}_{n}(\boldsymbol{\theta}),
\end{equation}
where $n$ is the excitation order of the excitation operator $\hat{T}_{n}$ and $\boldsymbol{\theta}$ are the circuit parameters. 
In this work, the coupled cluster expansion is truncated at single and double excitations and uses spin-adapted operators of the form\cite{Paldus1977,Piecuch1989,Packer1996}:
\begin{align}
    \ \hat{T}_{1}(\boldsymbol{\theta}) &= \sum_{ia}\theta_i^a \frac{1}{\sqrt{2}}\hat{E}_{ai} \label{eq:wf_par_1}\\
    \nonumber
    \ \hat{T}_{2}(\boldsymbol{\theta}) &= \sum_{i \geq j, a\geq b}\theta_{ij}^{ab} \frac{1}{2\sqrt{\left(1+\delta_{ab}\right)\left(1+\delta_{ij}\right)}}\left(\hat{E}_{ai}\hat{E}_{bj} + \hat{E}_{aj}\hat{E}_{bi}\right)\\
    &+  \sum_{i \geq j, a\geq b}\theta_{\phantom{'}ij}^{'ab} \frac{1}{2\sqrt{3}}\left(\hat{E}_{ai}\hat{E}_{bj} - \hat{E}_{aj}\hat{E}_{bi}\right)
    \label{eq:wf_par_2}
\end{align}
Here, $\hat{E}_{pq} = \hat{a}_{p,\alpha}^\dagger\hat{a}_{q,\alpha} + \hat{a}_{p,\beta}^\dagger\hat{a}_{q,\beta}$ is the singlet excitation operator. 
To find the parameters of the ground state, the electronic energy is minimized with respect to $\boldsymbol{\theta}$, which is obtained from the electronic Hamiltonian: 
%
\begin{equation}
    \hat{H} = \sum_{pq}h_{pq}\hat{E}_{pq} + \frac{1}{2}\sum_{pqrs}g_{pqrs}\hat{e}_{pqrs}~.
\end{equation}
Here, $\hat{e}_{pqrs} = \hat{E}_{pq}\hat{E}_{rs} - \delta_{qr}\hat{E}_{ps}$ is the two-electron excitation operator, $h_{pq}$ are one-electron molecular orbital integrals, and $g_{pqrs}$ are two-electron molecular orbital integrals.
In the VQE method\cite{peruzzo2014variational}, the circuit parameters are obtained by variational minimization of the energy
\begin{equation}
    E_\text{gs} = \min_{\boldsymbol{\theta}}\left<\Psi_{\mathrm{UCC}}\left(\boldsymbol{\theta}\right)\left|\hat{H}\right|\Psi_{\mathrm{UCC}}\left(\boldsymbol{\theta}\right)\right>.
\end{equation}

After a ground state circuit has been prepared and optimized, excitation energies and associated transition properties can be obtained with the quantum linear response approach. 
In qLR, excitation energies are obtained by classically solving a generalized eigenvalue problem that takes the form
\begin{equation}
    \mathbf{E}^{[2]} \mathbf{X}_k = E_{0k} \mathbf{S}^{[2]} \mathbf{X}_k~,
\end{equation}
where $k$ indexes each of the $N$ excitation energies.
The generalized eigenvalue problem has the following block structure
\begin{equation}
 \left(\begin{array}{cc}
\mathbf{A} & \mathbf{B}\\
\mathbf{B}^{*} & \mathbf{A}^{*}
\end{array}\right)\left(\begin{array}{c}
\mathbf{Y}_{k}\\
\mathbf{Z}_{k}
\end{array}\right)=E_{0k}\left(\begin{array}{cc}
\mathbf{\Sigma} & \mathbf{\Delta}\\
-\mathbf{\Delta}^{*} & \mathbf{-\Sigma}^{*}
\end{array}\right)\left(\begin{array}{c}
\mathbf{Y}_{k}\\
\mathbf{Z}_{k}
\end{array}\right).
\label{eq:qLR}
\end{equation}
The individual blocks contain elements defined as
\begin{equation}
    A_{ij}  = \left<\Psi\left| \comm{\hat{G}_i ^{\dagger}}{\comm{\hat{H}}{\hat{G}_j}}\right|\Psi\right>~, 
\end{equation}
\begin{equation}
    B_{ij}  = \left<\Psi\left| \comm{\hat{G}_i}{\comm{\hat{H}}{\hat{G}_j}}\right|\Psi\right>~,  
\end{equation}
\begin{equation}
    \Sigma_{ij}  = \left<\Psi\left| \comm{\hat{G}_i ^{\dagger}}{\hat{G}_j}\right|\Psi\right>~, 
\end{equation}
\begin{equation}
    \Delta_{ij}  = \left<\Psi\left| \comm{\hat{G}_i}{\hat{G}_j}\right|\Psi\right>~. 
\end{equation}

The $\hat{G}$ operators are the excitation operators, which in this work are spin-adapted just as the singles and doubles operators in the wavefunction parameterization.
This parameterization restricts the calculated excitation energies to be singlet excitations and reduces the dimensionality of the generalized eigenvalue problem compared to using spin-conserving operators.
%


By replacing the naive excitation operators, $\hat{G}$, with their self-consistent manifold 
\begin{align}
    \hat{G}^\mathrm{sc} = \boldsymbol{U} \hat{G} \boldsymbol{U}^\dagger
\end{align}
the overlap matrix becomes diagonal ($\mathbf{\Delta} = \mathbf{0}$ and $\mathbf{\Sigma}=\mathbf{I}$), and the off-diagonal blocks of the Hessian become zero ($\mathbf{B} = \mathbf{0}$). This is known as q-sc-EOM or q-sc-LR\cite{asthana2023quantum,kumar2023quantum} and simplifies Eq.~\eqref{eq:qLR} to a regular eigenvalue problem
\begin{equation}
    \mathbf{A}\mathbf{Y}_{k}=E_{0k}\mathbf{Y}_{k}.
    \label{eq:eigenvalue}
\end{equation}
The matrix elements of the eigenvalue problem are evaluated using a quantum computer.
As written, one needs to measure every element of the $\mathbf{A}$ matrix, after which the eigenvalues can be obtained by classical diagonalization.
However, when only a few eigenvalues (and associated eigenvectors) of Eq.~\eqref{eq:eigenvalue} are required, evaluating the complete matrix is not strictly necessary. 
Instead, eigenvalues can be obtained with Davidson methods\cite{DAVIDSON197587,liu1978simultaneous,olsen1988solution}, which only entail the action of Hessian-vector multiplication for arbitrary trial vectors. Such methods are widespread\cite{olsen1988solution} in classical quantum chemistry for obtaining excitation energies and solving linear response equations. 

The Hessian-vector product $\boldsymbol{\sigma}_\mathbf{b} = \mathbf{A}\mathbf{b}$ for a trial vector $\mathbf{b}$ can be approximated using the gradient $\mathbf{g}$ in a central finite-difference approach as\cite{chan1984nonlinearly}
\begin{equation}
     \boldsymbol{\sigma}_\mathbf{b} = \mathbf{A} \mathbf{b} 
        \approx  \frac{\mathbf{g}\left(\boldsymbol{\theta}, + h\mathbf{b}\right) - \mathbf{g}\left(\boldsymbol{\theta}, - h\mathbf{b}\right) }{2h} , 
        \label{eq:hvp}
    \end{equation}
where $h$ is a small positive parameter.
The linear transformation in Eq.~\eqref{eq:hvp} is a central part of matrix-free algorithms that can extract a few selected excitation energies of a molecule. 
To perform the Hessian-vector multiplication in Eq.~\eqref{eq:hvp}, we compute gradients of the excitation parameters for the states  $\left|\Psi(\boldsymbol{\theta},h\mathbf{b})\right>$ and $\left|\Psi(\boldsymbol{\theta},-h\mathbf{b})\right>$.
Explicitly, we compute gradients in the $\mathbf{\Theta}$ parameters on states that are defined as
\begin{equation}
    \left|\Psi(\boldsymbol{\theta},\mathbf{\Theta})\right> = U(\boldsymbol{\theta})U(\mathbf{\Theta})\left|0\right>,
\end{equation}
with $\mathbf{{\Theta}}=\pm h \mathbf{b}$.
On noiseless simulators, such a finite difference approach can work without issues for small values of the displacement parameter $h$. Realizing more efficient or noise-resilient implementations, e.g., through clever contractions of the trial vectors or deriving parameter-shift rules, will be part of future research.
We note that a similar approach was proposed by \citeauthor{parrish2021analytical} to compute state-averaged VQE (SA-VQE) Hessian-vector products\cite{parrish2021analytical}.

\subsection{The Davidson algorithm for excitation energies}\label{sec:theory_excitation_energy}
The Davidson method solves the eigenvalue problem 
(Eq.~\eqref{eq:eigenvalue}) using only Hessian-vector transformations. 
In this method, the $\mathbf{A}$ matrix is projected onto a subspace $\mathcal{B}$. The projected Hessian is diagonalized, and the subspace is expanded iteratively until convergence.
The goal of the algorithm is to extract a small number ($K$) of eigenvalues and associated eigenvectors of the full eigenvalue problem.

An initial set of normalized guess vectors $\mathcal{B} = \{\mathbf{b}_1,\mathbf{b}_2,\cdots,\mathbf{b}_L\}$ with $L \leq N$ is generated. The subspace dimension ($L$) is typically much smaller than the full-space dimension ($N$).
Our implementation generates initial guess vectors based on orbital energy differences\cite{parrish2016balancing}.
The Hessian-vector products $\boldsymbol{\sigma}_\mathbf{b}$ are then computed for each vector $\mathbf{b}$ in $\mathbf{\mathcal{B}}$ using Eq.~\eqref{eq:hvp} and are collected in the matrix $\boldsymbol{\sigma}_\mathcal{B}$.
Next, the projected Hessian matrix is evaluated as
\begin{equation}
    \mathbf{A}_\mathcal{B} = \mathcal{B}^{\dagger}\mathbf{A} \mathcal{B} = \mathcal{B}^{\dagger}\boldsymbol{\sigma}_\mathcal{B},
\end{equation}
As the dimension of $\mathbf{A}_\mathcal{B}$ is much smaller than that of the full Hessian, it is readily diagonalized using classical linear algebra routines to obtain a set of eigenvectors $\mathbf{z}_k$ (collected in the matrix $\mathbf{Z}$) and eigenvalues $\lambda_k$, of which only the $K$ lowest are kept. 
The (approximate) eigenvector matrix $\mathbf{X}$ of the excitation vectors are followingly constructed as 
\begin{equation}
    \mathbf{X} = \mathcal{B}\mathbf{Z}
    \label{eq:X=VZ}
\end{equation}

Next, residuals for the $K$ desired roots are computed as
\begin{align}
    \mathbf{r}_k &= \left(\mathbf{A} - \lambda_k \mathbf{I} \right) \mathbf{X}_{k} \\\nonumber
                 &= \mathbf{A}  \mathcal{B}\mathbf{Z}_k - \lambda_k \mathbf{I} \mathbf{X}_{k} \\\nonumber
                 &= \boldsymbol{\sigma}_{\mathcal{B}} \mathbf{Z}_k - \lambda_k \mathbf{I} \mathbf{X}_{k}
\end{align}
If the norm of the residual vectors is below some pre-defined threshold, the algorithm terminates, yielding $\lambda_k$ as the eigenvalues and $\mathbf{X}_{k}$ as the corresponding eigenvectors. 
Otherwise, correction vectors are computed as 
\begin{equation}
    \delta_{ki}  = \left(\lambda_k - \mathbf{A}_{ii} \right)^{-1} r_{ki}  
\end{equation}
In the equation above, $\mathbf{A}_{ii}$ is a diagonal element of the Hessian or some suitable approximation. We use a simple CI-inspired approximation using orbital energy differences in our implementation. If the norm of a correction vector is above some pre-defined threshold, the vector is normalized, orthogonalized to the other trial vectors, and appended to $\mathcal{B}$.
%
%
%
 
\subsection{Linear response with the Davidson method}\label{sec:theory_lr}
In frequency-dependent linear response theory with self-consistent operators, the following equation for a perturbation $B$ with frequency $\omega$ is solved

\begin{equation}
    \left(\mathbf{E}^{[2]} - \omega \mathbf{S}^{[2]} \right) \mathbf{X}_B(\omega) = \mathbf{V}_B .
    \label{eq:lin_resp_eq}
\end{equation}
Here, $\mathbf{V}_B$ is the property gradient, $\omega$ is the frequency, and the elements of the Hessian and generalized overlap matrices are defined as in 
Eq.~\eqref{eq:qLR}.
An element of property gradient can be evaluated as a derivative
\begin{equation}
    \mathbf{V}_{B,i} = \left<\Psi\left|\left[ \hat{B}, \hat{G}_i\right]\right|\Psi\right> = \frac{\partial}{\partial G_i} \left<\Psi(\theta, 0) \left|\hat{B}\right|\Psi(\theta, 0)\right>  
\end{equation}
When applying self-consistent operators, the full linear system of Eq.~\eqref{eq:lin_resp_eq} reduces to two independent 
(half-dimension) equations
\begin{equation}
    \left(\mathbf{A} - \omega \mathbf{I}\right) \mathbf{Y}_B (\omega)= \mathbf{V}_B,
\end{equation}
\begin{equation}
    \left(\mathbf{A} + \omega \mathbf{I}\right) \mathbf{Z}_B (\omega)= -\mathbf{V}_B^{*}.
\end{equation}
Solving these response equations gives response vectors $\mathbf{Y}_B$ and $\mathbf{Z}_B$ which can be used to compute linear response properties as
\begin{equation}
    \left<\left<A;B\right>\right>_\omega = \mathbf{V}_A \cdot \mathbf{Y}_B(\omega)  + \mathbf{V}_A^{*} \cdot \mathbf{Z}_B (\omega).
    \label{eq:linear_response_function}
\end{equation}
Static properties are recovered in the limit $\omega \rightarrow 0$.

The Davidson method can be straightforwardly generalized from solving the eigenvalue problem (Eq.~\eqref{eq:qLR}) to solving the linear equation in Eq.~\eqref{eq:lin_resp_eq}.
We use the same setup of defining a subspace $\mathcal{B}$, onto which the linear response equation is projected
\begin{equation}
        \left(\mathbf{A}_{\mathcal{B}} - \omega \mathbf{I} \right) \mathbf{X}_{B,\mathcal{B}}(\omega) = \mathbf{V}_{B,{\mathcal{B}}}
        \label{eq:lin_resp_eq_projected}.
\end{equation}
Here, the projected Hessian and property vector are defined as
\begin{align}
        \mathbf{A}_\mathcal{B} = \mathcal{B}^{\dagger}\mathbf{A} \mathcal{B} \\
        \mathbf{V}_{B,\mathcal{B}} = \mathcal{B}^{\dagger}\mathbf{V}_B 
\end{align}

The solution vector can be lifted from the subspace to the full space as
\begin{equation}
    \mathbf{X}_B(\omega) = \mathcal{B} \mathbf{X}_{B,\mathcal{B}}(\omega) .
\end{equation}

The general flow of the linear response Davidson routine is almost identical to the eigenvalue case. Given an initial subspace $\mathcal{B}$, the projected linear equation (Eq.~\eqref{eq:lin_resp_eq_projected}) is solved to obtain the subspace solution vector $\mathbf{X}_{B,\mathcal{B}}(\omega)$. The residual is then computed as
\begin{align}
\nonumber
    \mathbf{r} &= \mathbf{V}_B - (\mathbf{A} - \omega \mathbf{I}) \mathbf{X}_{B}  \\
    \nonumber
    &= \mathbf{V}_B - (\mathbf{A} - \omega \mathbf{I}) \mathcal{B}\mathbf{X}_{B,\mathcal{B}}  \\
    \nonumber
    &= \mathbf{V}_B - (\mathbf{A}\mathcal{B}\mathbf{X}_{B,\mathcal{B}} - \omega \mathbf{I}\mathcal{B}\mathbf{X}_{B,\mathcal{B}}) \\
    &= \mathbf{V}_B - (\sigma_\mathcal{B}\mathbf{X}_{B,\mathcal{B}} - \omega \mathbf{I}\mathbf{X}_{B})
\end{align}
The solution vector is returned if the residual norm is below some tolerance. Otherwise, a new trial vector is added to the subspace from the preconditioned residual correction vector
\begin{equation}
\label{precondLRE}
    \delta_{i} = \left(\mathbf{A}_{ii} - \omega \right)^{-1} r_{i}
\end{equation}
which is normalized, orthogonalized to $\mathcal{B}$, and appended to the subspace vectors.
When applying the self-consistent operator manifold, we note that the generalized overlap matrix is a unit matrix, while in a more general case, one would also need to project the generalized overlap matrix. One of the major ingredients in the Davidson solver is, therefore, still the computation of Hessian-vector products with arbitrary trial vectors, for which the approach in Eq.~\eqref{eq:hvp} remains applicable.
The gradient vector, preconditioned as in  Eq.~\eqref{precondLRE}, is often used as a start vector in the iterative procedure to solve the non-homogeneous linear response equations.

Linear response functions can be used to obtain many molecular properties.
In this work, we focus on the polarizability, which is equal to (minus) the dipole--dipole response function
\begin{equation}
    \alpha_{ij}(\omega) = -\left<\left<\hat{\mu}_i;\hat{\mu}_j\right>\right>_\omega ,
\end{equation}
with $i$ and $j$ referring to cartesian components.

\subsection{Damped response}\label{sec:theory_damped}
The linear response function in Eq.~\eqref{eq:linear_response_function} has poles (divergences) at frequencies corresponding to excitation energies, which can easily be seen from the spectral representation of linear response properties
\begin{equation}
        \left<\left<A;B\right>\right>_\omega = -\sum_{k\neq 0}\left(\frac{\left<0\left|\hat{A}\right|k\right>\left<k\left|\hat{B}\right|0\right>}{\omega_k-\omega} +\frac{\left<0\left|\hat{B}\right|k\right>\left<k\left|\hat{A}\right|0\right>}{\omega_k+\omega}\right)~, 
        \label{eq:sos_polarizability}
\end{equation}
where divergences happen when $\omega = \pm \omega_k$.
Linear response theory can also be formulated in a framework that is convergent at all frequencies, with the approach known as resonant-convergent damped response theory, also known as the complex polarization propagator.~\cite{norman:2001cpplinear,norman:2005nonlinear,kristensen2009quasidampedresponse,norman:2011perspective,helgaker:2012response,kauczor2013dampedcc,faber2019:rixs}
In damped response theory, an imaginary damping parameter ($i\gamma$) is introduced, and the polarizabilities become complex-valued but finite at all frequencies. 
In the sum-over-states representation, the introduction of $\gamma$ leads to
\begin{equation}
        \left<\left<A;B\right>\right>_{\omega,\gamma} = -\sum_{k\neq 0}\left(\frac{\left<0\left|\hat{A}\right|k\right>\left<k\left|\hat{B}\right|0\right>}{\omega_k-(\omega+i\gamma)} +\frac{\left<0\left|\hat{B}\right|k\right>\left<k\left|\hat{A}\right|0\right>}{\omega_k+(\omega+i\gamma)}\right)~,
\end{equation}
which remains finite even at $\omega=\pm\omega_k$.
Practically speaking, the extension of regular frequency-dependent linear response theory to damped response theory entails replacing the real frequency $\omega$ with $\omega + i\gamma$, where $\gamma$ is a real parameter.
The classical parts of the Davidson method can be carried out without issue in complex algebra, with the only requirement that transposes are replaced by the complex adjoint.
This approach leads to complex trial vectors, which can be decomposed into purely real and imaginary parts as
\begin{equation}
    \mathbf{b} = \mathbf{w} + i\mathbf{u}. \label{eq:complex_trial_vector}
\end{equation}
A Hessian-vector multiplication with a complex trial vector can then be computed as
\begin{equation}
    \boldsymbol{\sigma}_\mathbf{b} = \boldsymbol{\sigma}_\mathbf{w} + i \boldsymbol{\sigma}_\mathbf{u}.
\end{equation}
The extension to damped response does not lead to any new types of ``quantum'' operations since the standard Hessian-vector multiplication can be applied.
We note that as an alternative to complex algebra, one could also decompose the solution vector as in Eq.~\eqref{eq:complex_trial_vector} and introduce this decomposition in the linear response equations, which allows the problem to be formulated as a regular (real) problem with twice the size.
%
%

%

With a complex linear response code, several interesting properties can be extracted.
Absorption cross-sections can be extracted from the imaginary part of the complex dipole--dipole polarizability as\cite{fransson2016k}

\begin{equation}
    \sigma(\omega) = \frac{\omega}{\epsilon_0 c} \mathrm{Im}(\alpha_{\mathrm{iso}}(\omega)),
    \label{eq:sigma_abs}
\end{equation}
where the isotropic polarizability is the average of the diagonal components, $\alpha_\mathrm{iso} = \frac{1}{3} (\alpha_{xx}  + \alpha_{yy} + \alpha_{zz})$.

Several other molecular properties can also be extracted. For example, long-range orientation-averaged dipole-dipole dispersion $C_6$ coefficients between two systems, $A$ and $B$, can be evaluated using the Casimir-Polder relation\cite{casimir1948influence,kauczor2013non}

\begin{equation}
    C^{AB}_6 = \frac{3 \hbar}{\pi} \int_0^\infty  \alpha_{\mathrm{iso}}^{A}(i\omega) \alpha_{\mathrm{iso}}^{B}(i\omega) \mathrm{d}\omega,
    \label{eq:C6}
\end{equation}
where $ \alpha_\mathrm{iso}^{A}(i\omega)$ is the isotropic polarizability of system $A$ evaluated at the purely imaginary frequency $i\omega$.
Following the procedure outlined in Ref. \citenum{kauczor2013non}, the integration in Eq.~\eqref{eq:C6} can be performed using Gauss-Legendre quadrature by mapping the $[0,\infty]$ range to $[-1,1]$ via a change of variables
\begin{equation}
    \omega = \omega_0 \left(\frac{1-t}{1+t}\right)
\end{equation}
where an appropriate value of $\omega_0$ is 0.3\cite{kauczor2013non}. 
The interval transformation gives the Casimir-Polder integral as
\begin{equation}
    C^{AB}_6 = \frac{3\hbar}{\pi} \int_{-1}^{1} \frac{2\omega_0}{(1+t)^2}  \alpha_\mathrm{iso}^{A}(i\omega(t)) \alpha_\mathrm{iso}^{B}(i\omega(t)) \mathrm{d}t
\end{equation}
The integral is then evaluated with Gauss-Legendre quadrature as
\begin{equation}
    C^{AB}_6 \approx \frac{3 \hbar}{\pi} \sum_k w_k \frac{2\omega_0}{(1+t_k)^2}  \left\{\alpha_\mathrm{iso}^{A}(i\omega(t_k)) \alpha_\mathrm{iso}^{B}(i\omega(t_k)) \right\},
\end{equation}
where $w_k$ and $t_k$ are quadrature nodes and weights, respectively, and $\omega(t_k) = \omega_0 (1-t_k)/(1+t_k)$.
We use a 12-point Gauss-Legendre quadrature to approximate the integral.

\section{Computational Details}\label{sec:computational_details}

Excitation energies of hydrogen chains in a ladder-like configuration (see Figure \ref{fig:hn_chain}) were computed with an STO-3G\cite{hehre1969a,hehre1970a} basis with UCCSD.
We modified the existing UCCSD implementation from Pennylane\cite{pennylane} (version 0.34.0) to support spin-adapted (singlet) fermionic excitation operators and to enable the computation of gradients of excitation operators.
Our code is publically available on Github\cite{subspaceresponsegit}.
The ground-state UCCSD wave functions were optimized using the SLSQP optimizer\cite{slsqp} from SciPy\cite{virtanen2020scipy}. Quantum circuits were simulated using the Pennylane Lightning state-vector simulator, with gradients evaluated using the adjoint differentiation\cite{jones2020efficient} approach, unless otherwise noted. Hessian-vector products (Eq.~\eqref{eq:hvp}) were performed with a step size $h=10^{-6}$ for the noise-free simulations.
One- and two-electron integrals and Hartree-Fock wave functions were obtained with PySCF\cite{pyscf}.
FCI calculations of the excitation energies were carried out using the Dalton program~\cite{daltonpaper}.
All calculations were carried out on a desktop computer equipped with an Intel i9-7980XE CPU.

Geometries of \ce{LiH}, \ce{OH-}, and \ce{H2O} were optimized at the FCI/6-31G\cite{dill1975a,ditchfield1971a,hehre1972a} level of theory using the Dalton program\cite{daltonpaper}.
The geometry of \ce{NH3} was obtained from MP2/cc-pVTZ\cite{dunning1989a} calculations with the Orca program\cite{orca}.
All geometries are available in the supporting information.
UCCSD calculations of the polarizabilities were carried out with the same settings as described above. All calculations were carried out as full-space calculations, i.e., we did not apply any frozen-core approximation. The CCSD and FCI polarizabilities were computed with the Dalton program. A development version of Dalton~\cite{kauczor2013dampedcc} was used for the damped response CCSD calculations of the polarizability at imaginary frequencies used to obtain the $C_6$ coefficients.

\section{Results and Discussion}



\subsection{Excitation energies}\label{sec:results_excitation_energy}
\begin{figure}
    \centering
    \includegraphics[width=5cm]{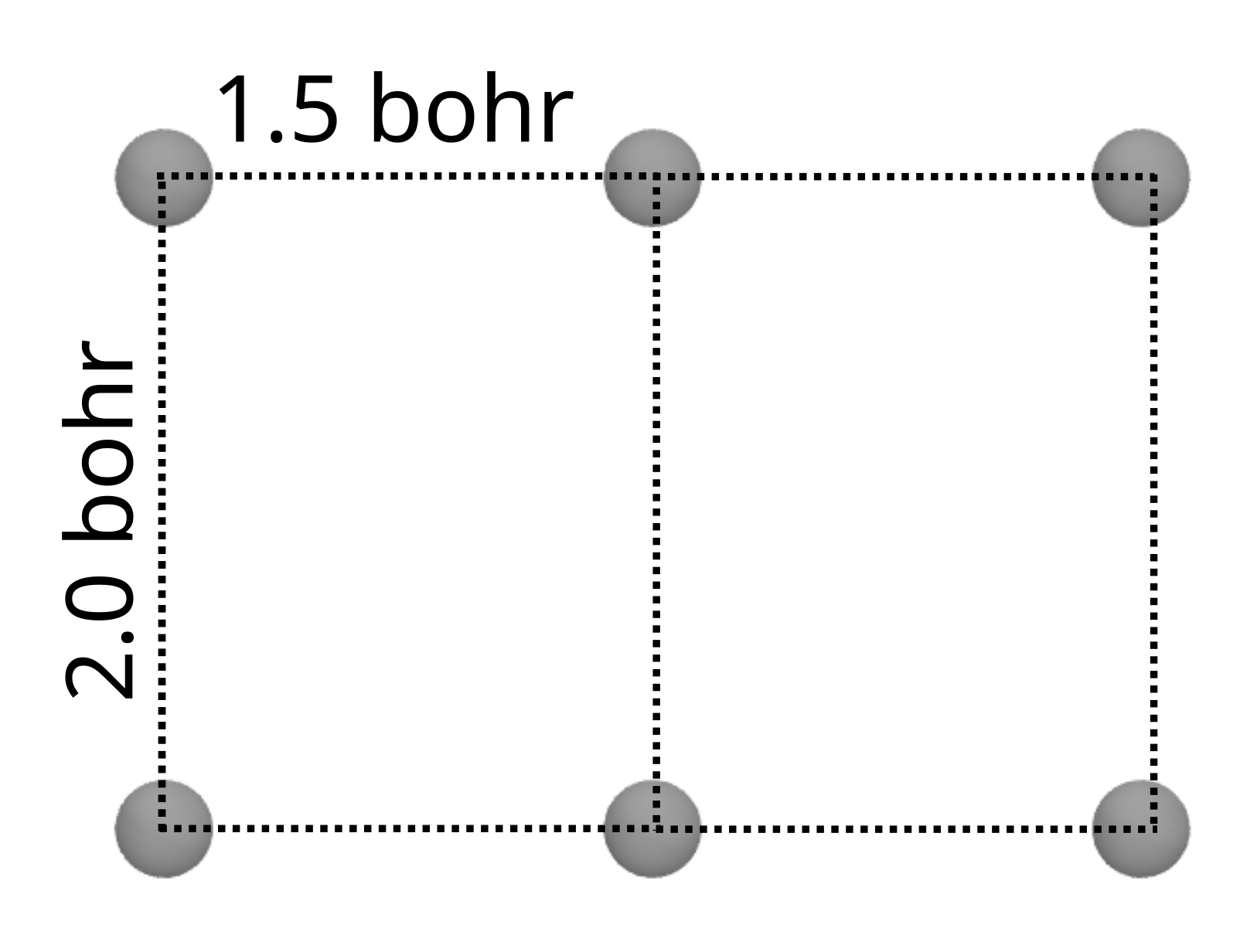}
    \caption{The geometry of the H$_{2n}$ chains, shown for $2n=6$.}
    \label{fig:hn_chain}
\end{figure}

Table \ref{tab:hn_chain} shows the lowest excitation energy and associated timings on H${_{2n}}$ hydrogen-ladder chains, using either FCI, UCCSD with explicit construction of the Hessian, or UCCSD using the Davidson solver. The dimension of the subspace for the converged excitation energies is also reported.
For the subspace approaches, the two lowest roots were converged.
The geometry of the hydrogen chains is sketched in Figure \ref{fig:hn_chain}.

The lowest excitation energy varies considerably and somewhat unpredictably with the size of the hydrogen chain. For example, the excitation energy for H$_6$ is 14.14 eV, which suddenly drops to 2.98 eV with H$_8$, only to return to 10.54 eV for H$_{10}$. 
The UCCSD excitation energies computed with the explicit construction or the Davidson solver are identical to convergence thresholds for all the considered excitations.
The UCCSD excitation energies generally agree well with the FCI results, with an exact or near-exact agreement for the smaller hydrogen chains. The largest error relative to FCI (0.37 eV deviation) occurs for H$_8$, but the trend in the excitation energy is well reproduced.
Regarding timings, the Davidson method offers clear improvements over the full diagonalization approach, with calculations that are an order of magnitude faster for the larger systems.
The speed-up in timings is almost entirely driven by the need to evaluate fewer Hessian-vector products. 
Formally, the full diagonalization approach can be seen as evaluating a Hessian-vector product for each basis vector of the matrix in Eq.~\eqref{eq:eigenvalue}.
When parametrizing the excitation operator manifold with single and double excitations, the matrix dimension grows asymptotically as $N^4$ with the system size. In contrast, the required dimensions of the UCCSD and FCI subspace appear to level out at a size of around 13--14. We note that the FCI calculations are carried out using the Dalton program, which uses a Davidson-like\cite{olsen1988solution} method to extract excitation energies.
For FCI, we report the dimension as half the subspace size since the FCI response solver in Dalton uses paired trial vectors.

\begin{table} 
    \centering
\begin{tabular}{c|ccc|cc|ccc}
 & \multicolumn{3}{c|}{$\Delta E_{\mathrm{exc}}$ (eV)} & \multicolumn{2}{c|}{Time (seconds)} & \multicolumn{3}{c}{Dimension}\tabularnewline
 & FCI  & UCCSD$^{a}$  & UCCSD$^{b}$  & UCCSD  & UCCSD & FCI  & UCCSD & UCCSD\tabularnewline
$2n$  &  & Full diag. & Davidson & Full diag. & Davidson &  & Full diag. & Davidson\tabularnewline
\hline 
2  & 19.44  & 19.44  & 19.44  & 0.0450  & 0.0431  & 2  & 2  & 2\tabularnewline
4  & 10.08  & 10.09  & 10.09  & 6.85  & 4.97  & 2  & 14  & 10\tabularnewline
6  & 14.14  & 14.19  & 14.19  & 196  & 43.46  & 8  & 54  & 12\tabularnewline
8  & 2.98  & 3.35  & 3.35  & 9930  & 950  & 13  & 152  & 14\tabularnewline
10  & 10.54  & -- $^{a}$  & 10.45  & --$^{a}$  & 56400  & 13  & 350  & 14 \tabularnewline
\end{tabular}
    \caption{Excitation energies, timings, and matrix dimensions of the (reduced) Hessian for $\mathrm{H}_{2n}$/STO-3G chains. For FCI, we report half the dimension of the reduced space. $^{a}$ Not determined. The calculation would take an estimated 16 days.}
    \label{tab:hn_chain}
\end{table}

\subsection{Sampling noise}\label{sec:results_noise}
Next, we consider the effects of sampling noise in determining excitation energies, using \ce{H2} as a minimal example.
For this set of tests, we used the parameter-shift rule for evaluating gradients in Eq.~\eqref{eq:hvp} since the adjoint differentiation approach does not permit the modeling of shot noise.
When using spin-adapted singlet excitation operators, the dimension of the Hessian is $2\times2$ for \ce{H2}/STO-3G, corresponding to one single and one double excitation.
In this small example, the Davidson method terminates trivially in a single iteration when we use the initial trial vectors $b_1 = (1, 0)$ and $b_2 = (0, 1)$ to compute the Hessian.
In the presence of sampling noise, a potentially important new complication emerges, namely that the hermiticity of the Hessian is destroyed.
We consider three options to tackle the non-hermiticity of $\mathbf{A}_\mathcal{B}$ in our algorithms: 1) using only the upper triangle of the Hessian (assigning $A_{\mathcal{B},ji} = A_{\mathcal{B},ij}$), 2) solving the eigenvalue with a non-hermitian Hessian, and 3) symmetrizing the Hessian as $\tilde{\mathbf{A}}_{\mathcal{B}} = \frac{1}{2} \left(\mathbf{A}_{\mathcal{B}} + \mathbf{A}_{\mathcal{B}}^{\dagger}\right)$.
Additionally, the finite difference approach used to form the Hessian-vector products (Eq.~\eqref{eq:hvp}) becomes sensitive to noise for small values of $h$.

\begin{figure}
    \centering
    \includegraphics[width=1.0\textwidth]{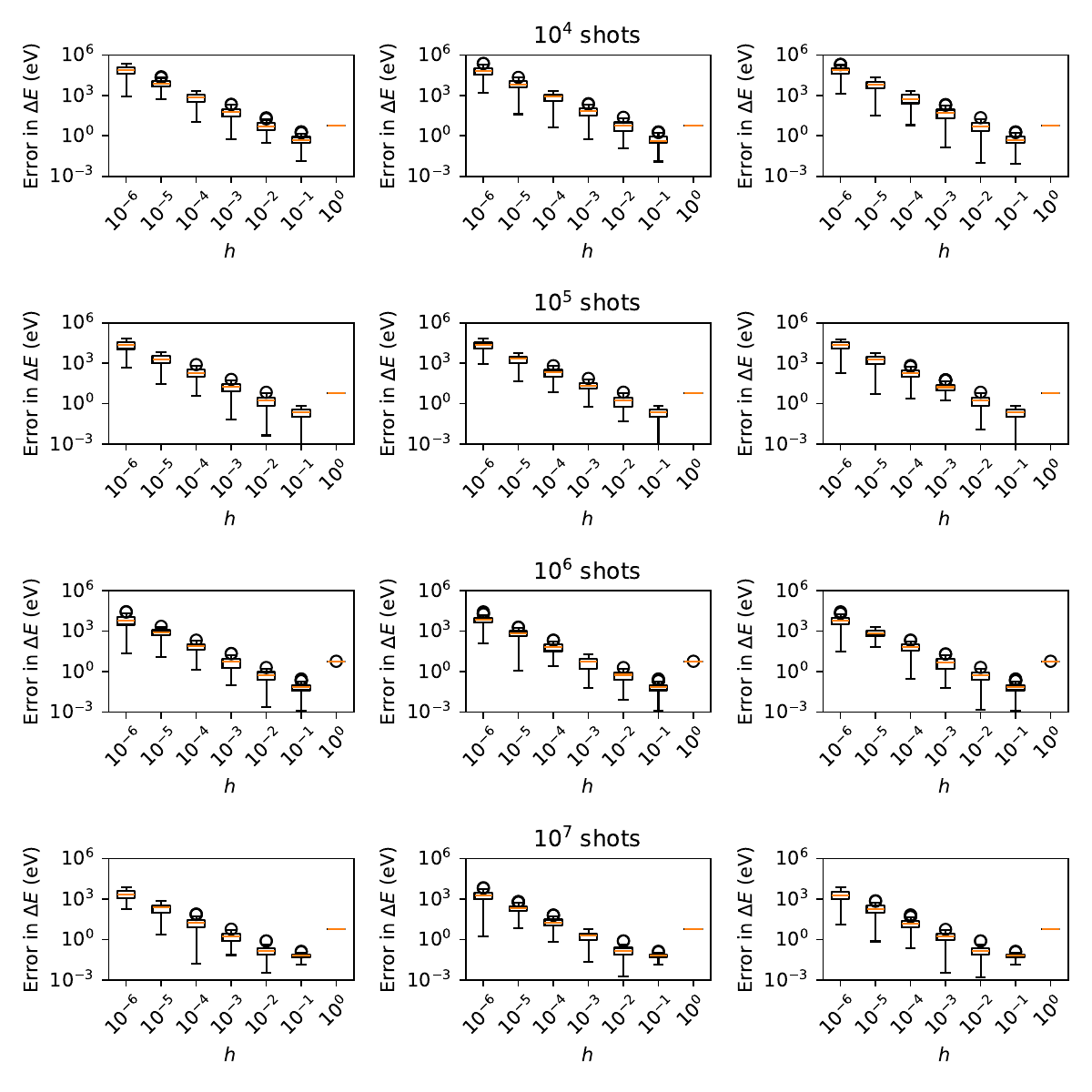}
    \caption{Absolute error in the lowest excitation energy of \ce{H2}/STO-3G for a fixed number of shots ($10^4$ -- $10^7$) as a function of the central difference step parameter ($h$). The distributions are taken across $N=100$ samples, shown as a box plot. The orange lines indicate the mean error. The noise-free result is used as a reference. For the non-hermiticity treatment, the left panels use the upper triangular part of the Hessian, the middle panels solve the non-hermitian problem, and the right panels use a symmetrized Hessian.}
    \label{fig:h-cal}
\end{figure}

We consider the effects of the hermiticity treatment and the size of the finite-difference step in Figure \ref{fig:h-cal}, where we have computed excitation energies with each of the three approaches to solving the eigenvalue problem for different values of the finite difference step parameter $h$, plotting the distribution $(N=100)$ for the absolute error in the lowest excitation energy.
For small values of $h$, the finite difference scheme is completely unreliable due to the sampling errors, leading to extremely large errors in the excitation energies.
Increasing $h$ initially decreases the error in the excitation energy until truncation errors in the finite difference scheme become significant. The optimal balance for every tested shot count amount appears surprisingly large at around $h=0.1$, where, with $10^7$ shots, the mean absolute error becomes 0.062 eV.
By increasing $h$ further (to the regime of truncation/discretization error), the sampling errors are almost completely avoided, as reflected by the hardly visible error bars at $h=1.0$. The truncation errors lead to significant misestimation of the parameter derivatives and, ultimately, excitation errors of more than 5 eV with $10^7$ shots.
We note that the optimal choice of $h$ and error characteristics will likely be system-dependent.
The tradeoff between sampling/truncation errors could be avoided if a parameter-shift rule was developed to evaluate the Hessian-vector product; however, as discussed by \citeauthor{parrish2021analytical}, the practical implementation of such a scheme is non-trivial\cite{parrish2021analytical}. 
The choice of restoring the Hessian's hermiticity turns out to be less consequential, with all three approaches having comparable performance.
When sampling is deeply insufficient (low $h$), the Hessian is so poorly described that all methods yield poor results. On the other hand, around the optimal $h$, the hermiticity violations become small, making all the approaches nearly equivalent.

\begin{figure}
    \centering
    \includegraphics{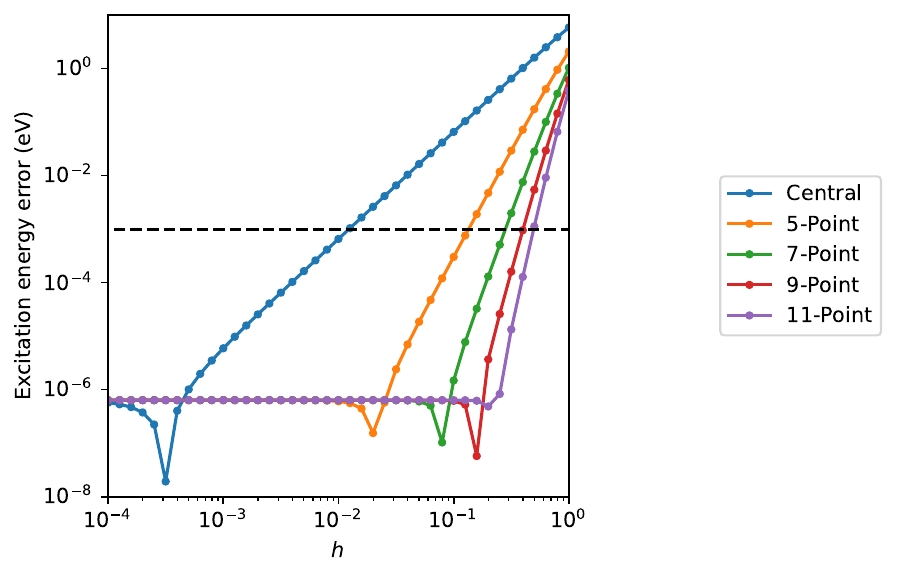}
    \caption{Absolute error in the lowest excitation energy of noiseless \ce{H2}/STO-3G as a function of the finite difference step parameter for various finite difference schemes. The black line indicates a target error of $10^{-3}$ eV.}
    \label{fig:h2_stencil_noisefree}
\end{figure}

From Figure \ref{fig:h-cal}, we find that using large step values $h$ is desirable since statistical noise errors can be suppressed. 
Therefore, we investigated whether more elaborate finite difference schemes could allow larger step sizes.
To examine this, we computed excitation energies of \ce{H2}  with higher-order numerical derivatives, namely 5-, 7-, 9, and 11-point symmetric stencils. 
We first consider the noise-free case, shown in Figure \ref{fig:h2_stencil_noisefree}.
All the finite difference schemes can achieve errors below $10^{-6}$~eV for the excitation energy with a proper choice of $h$. The region of stability widens when using the higher-order stencils. In particular, it is possible to use significantly larger step sizes without compromising the quality of the computed excitation energy.
Supposing we can accept an error of $10^{-3}$ eV in the excitation energy, the largest permissible step is around $h=0.012$ with the central difference scheme. The 5-point stencil allows a larger $h=0.13$, while the 7-, 9, and 11-point stencils allow up to $h=0.28$, $h=0.40$ and $h=0.50$. 
These larger step sizes are well into the region of significantly diminished statistical noise errors from Figure \ref{fig:h-cal}.

\begin{figure}
    \centering
    \includegraphics[width=1.0\textwidth]{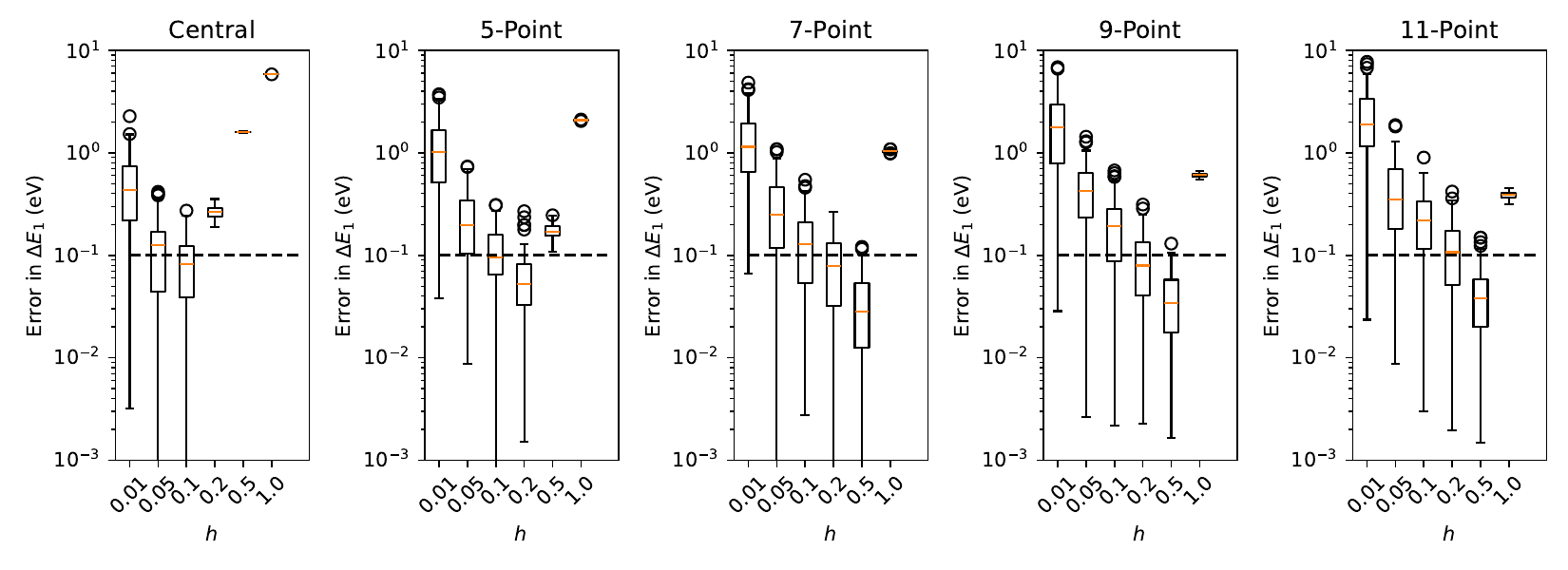}
    \caption{Absolute error in the lowest excitation energy of \ce{H2}/STO-3G for a fixed number of shots ($10^6$) as a function of the finite difference step parameter ($h$) for different finite difference schemes. The distributions are taken across $N=100$ samples. The orange lines indicate the mean error. The noise-free result is used as a reference. Each method is allowed a fixed shot budget: the 5, 7, 9, and 11-point schemes use $\frac{1}{2}$, $\frac{1}{3}$, $\frac{1}{4}$, and $\frac{1}{5}$ the amount of shots for each gradient evaluation compared to the central difference scheme.}
    \label{fig:h-cal-stencils}
\end{figure}

Next, we apply the higher-order finite difference stencils in a setting with sampling noise. The higher-order schemes require 4, 6, 8, and 10 gradient evaluations compared to just two for the central difference ("3-point") scheme.
Thus, we scale the shot count by the gradient evaluation count to provide fair comparisons for the shot count costs of the different finite difference approaches, i.e., we allow half the amount of shots per gradient evaluation for the 5-point scheme compared to the central difference scheme. 
As shown in Figure \ref{fig:h-cal-stencils}, the higher-order finite difference schemes allow for stability at significantly larger step parameters $h$ even in the presence of statistical noise.
The optimal $h$, in the sense of the lowest error in the excitation energy, gradually increases from $h=0.1$ with the central difference scheme to $h=0.5$ with the 9-point stencil.
On balance, this means that the mean absolute error in the excitation energy can be decreased by applying the higher-order difference schemes.
With a $10^6$ shot budget, the best mean absolute deviation decreases from $0.087$ eV with the central difference approach to $0.062$ and $0.036$ eV with the  5- and 7-point stencils, then increases slightly to $0.040$ and $0.044$ eV with the, 9- and 11-point stencils. Evidently, the increased statistical error due to the lower shot count in the higher-order finite-difference schemes becomes more significant than the associated decrease in truncation error.
For the present system, the optimal balance thus appears to be the 7-point stencil with a $h = 0.5$. Again, we note that, in general, the optimal balance will likely be somewhat system-dependent.

\subsection{Static polarizabilities}\label{sec:results_polarizability}
We now move on to consider the calculation of more general response properties (in a noise-free setting). Table \ref{tab:static_polarizability} shows static polarizabilities of selected small molecules computed using FCI, UCCSD, or conventional CCSD. 
All methods agree for the \ce{H2} molecule. As expected, the double excitations in UCCSD/CCSD recover the FCI solution for this two-electron system.
UCCSD and CCSD also have a good reproduction of the FCI polarizabilities of the equilibrium geometries of \ce{LiH}, water, the hydroxide ion, and ammonia.
The performance of UCCSD and CCSD is comparable for the equilibrium geometries, with mean absolute deviations of 0.0038 and 0.0033 a.u. for UCCSD and CCSD, respectively.
We also consider (symmetrically) stretched versions of these small molecules to generate systems with a more challenging electronic structure. These partially dissociated systems introduce stronger static correlation, which can be difficult to model. 
Nevertheless, we find that the stretched \ce{LiH} and \ce{OH-} systems can be described well by both UCCSD and CCSD. This can perhaps be rationalized from the observation that the dissociation in these systems involves just a single $\sigma$ bond (one pair of electrons). Since both methods include double excitations, they can describe this single bond-breaking correctly.
On the other hand, the stretched water and ammonia systems prove more challenging. UCCSD manages a mean absolute deviation (MAD) of 0.014 a.u. for water, which is considerably better than the CCSD result (0.84 a.u.).
The stretched ammonia system is even more challenging, with the CCSD polarizabilities being qualitatively wrong, having two \emph{negative} components in the ground-state polarizability. The UCCSD polarizability is qualitatively correct for this system, managing a MAD of 0.21 a.u. 
The exact ground-state polarizability is always non-negative, as seen from the sum-over-states expression of the polarizability (see Eq.~\eqref{eq:sos_polarizability}) since both the denominators (excitation energies) and numerators (squared transition moments) are positive. Excited-state polarizabilities can be negative. This could indicate that the CCSD wave function has converged to an excited state. However, we did not find any negative excitation energies with CCSD, although complex eigenvalues appeared in the solution of the CCSD linear response equations. Additionally, we also characterized the stability of the HF ground-state wave function, which revealed an instability towards a UHF solution with several negative eigenvalues, i.e., a triplet instability. 

\begin{table} 
    \centering
\begin{tabular}{l|rrr|rrr|rrr}
 & \multicolumn{3}{c|}{FCI} & \multicolumn{3}{c|}{UCCSD} & \multicolumn{3}{c}{CCSD}\tabularnewline
 & $\alpha_{xx}$ & $\alpha_{yy}$ & $\alpha_{zz}$ & $\alpha_{xx}$ & $\alpha_{yy}$ & $\alpha_{zz}$ & $\alpha_{xx}$ & $\alpha_{yy}$ & $\alpha_{zz}$\tabularnewline
\hline 
$\ce{H2}$ & 0.000 & 0.000 & 2.775 & 0.000 & 0.000 & 2.775 & 0.000 & 0.000 & 2.775\tabularnewline
$\ce{LiH}$ & 22.801 & 22.801 & 12.400 & 22.788 & 22.788 & 12.377 & 22.793 & 22.793 & 12.387\tabularnewline
$\ce{LiH}$ (2$R_{e}$) & 69.779 & 69.779 & 98.258 & 69.841 & 69.841 & 98.385 & 69.821 & 69.821 & 98.281\tabularnewline
$\ce{H2O}$ & 0.056 & 5.395 & 2.000 & 0.056 & 5.393 & 2.000 & 0.056 & 5.399 & 1.992\tabularnewline
$\ce{H2O}$ (2$R_{e}$) & 0.012 & 5.207 & 2.858 & 0.012 & 5.191 & 2.832 & 0.007 & 3.216 & 2.341\tabularnewline
$\ce{OH-}$ & 0.035 & 0.035 & 4.325 & 0.035 & 0.035 & 4.325 & 0.035 & 0.035 & 4.325\tabularnewline
$\ce{OH-}$ (2$R_{e}$) & 0.054 & 0.054 & 37.421 & 0.054 & 0.054 & 37.421 & 0.054 & 0.054 & 37.421\tabularnewline
$\ce{NH3}$ & 4.980 & 4.980 & 1.160 & 4.977 & 4.977 & 1.158 & 4.979 & 4.979 & 1.152\tabularnewline
$\ce{NH3}$ (2$R_{e}$) & 7.349 & 7.349 & 3.755 & 7.655 & 7.606 & 3.690 & -0.935 & -0.935 & 2.388\tabularnewline
\end{tabular}
    \caption{Polarizabilities (in a.u.), for selected molecules (STO-3G basis) computed with FCI, UCCSD, or CCSD.}
    \label{tab:static_polarizability}
\end{table}


\subsection{Dynamic polarizabilities}\label{sec:results_dynamic_polarizability}
\begin{figure}
    \centering
    \includegraphics{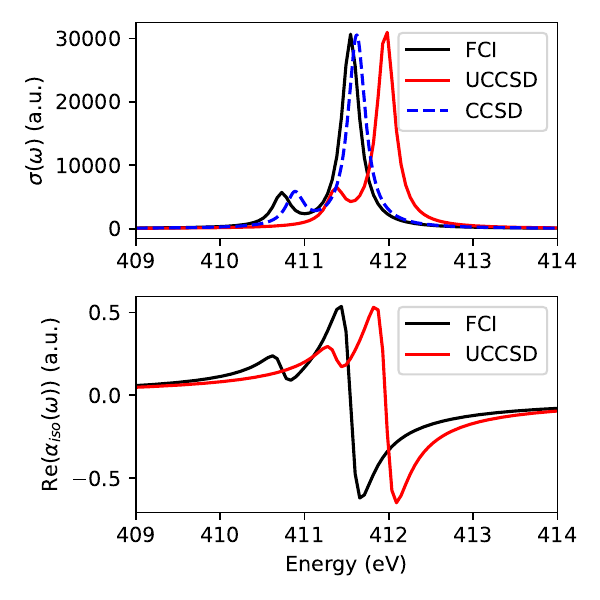}
    \caption{Imaginary and real parts of the isotropic polarizability of \ce{NH3}/STO-3G near the nitrogen K-edge X-ray absorption energy region. For the imaginary part, we plot the absorption cross-section (see Eq.~\eqref{eq:sigma_abs}). The spectra are computed using damped response theory with FCI or UCCSD. We also show CCSD spectra generated from excitation energies and oscillator strengths with a Lorentzian broadening.}
    \label{fig:nh3-sto3g-damped}
\end{figure}

Next, we turn to compute resonant-convergent frequency-dependent molecular properties with damped response theory. Figure \ref{fig:nh3-sto3g-damped} shows the imaginary (corresponding to absorption) and real (corresponding to dispersion) parts of the damped dipole--dipole polarizability of ammonia (in a STO-3G basis).
The spectra were computed in the energy range corresponding to the nitrogen K-edge X-ray absorption.
We used a damping factor $\gamma=0.004556$ a.u. (1000 $\mathrm{cm}^{-1}$).
The UCCSD absorption spectrum matches the FCI result qualitatively well, with a lower-intensity pre-edge feature followed by a more intense peak.
With UCCSD, the intense peak is blue-shifted by about 0.4 eV, while the lower-intensity peak is shifted by about 0.6 eV.
Due to the rather small basis set employed, any comparisons to experiments are not necessarily meaningful. However, the transitions at least occur in the correct general region for nitrogen K-edge X-ray absorption.
We should also point out that the density of states in the region is quite small when employing such a minimal basis set expansion, with only three excited states within the plotted energy region. Thus, it would likely be more computationally efficient to compute (targeted) excitation energies and associated oscillator strengths to obtain the spectrum.
We also show spectra generated from a Lorentzian convolution of CCSD excitation energies and oscillator strengths. As it turns out, the conventional CCSD spectra perform better than the UCCSD ones, with a smaller shift of 0.08 eV relative to the FCI spectrum. Clearly, one should not always expect UCCSD to outperform classical CCSD. However, ammonia has a relatively simple electronic structure at the equilibrium geometry, and as shown with the static polarizabilities in Table \ref{tab:static_polarizability}, the UCCSD method works well in more complicated systems, including systems with multiconfigurational character.

\subsection{Dispersion coefficients}\label{sec:results_c6}
Table \ref{tab:c6-coeffs} shows computed $C_6$ coefficients for the same set of molecules used in Table \ref{tab:static_polarizability} with the same STO-3G basis.
We compare $C_6$ coefficients computed with UCCSD to coefficients evaluated with an FCI wave function.
The $C_6$ coefficients vary considerably, from relatively small in magnitude (0.62 a.u. for \ce{H2}) to much more substantial values (409 a.u. for stretched \ce{LiH}).
Nevertheless, the UCCSD $C_6$ coefficients generally reproduce the reference FCI results very well, with relative errors below 0.1\% at equilibrium geometries.
The stretched geometries are more challenging, particularly for water and ammonia, where we find relative errors of up to 9\%. These systems were also challenging cases for the static polarizabilities.
The CCSD $C_6$ coefficients are reproduced well for the equilibrium geometries, but the method fails to give even qualitatively accurate results for the stretched water and ammonia systems, where the $C_6$ coefficients are either strongly overestimated (water, $+$55\%) or underestimated (ammonia, $-$99.9\%). 
Thus, while CCSD might give a good description of the dispersion coefficients for the equilibrium geometries, the method gives qualitatively incorrect results for some of the stretched systems.
Meanwhile, the UCCSD parametrization works well across both equilibrium and stretched geometries since it is able to handle systems with strong static correlation or multiconfigurational character.

\begin{table}
\centering %
\begin{tabular}{l|r|rr|rr}
 & FCI  & \multicolumn{2}{c|}{UCCSD}  & \multicolumn{2}{c}{CCSD} \tabularnewline
\hline 
$\ce{H2}$  & 0.62  & 0.62  & (0.00) &  0.62 &  (0.00)\tabularnewline
$\ce{LiH}$  & 50.84  & 50.79  & ($-$0.09) & 50.81 & ($-$0.05) \tabularnewline
$\ce{LiH}$ (2$R_{e}$)  & 408.71  & 409.24  & (0.13) & 409.27 & (0.14) \tabularnewline
$\ce{H2O}$  & 3.79  & 3.79  & ($-$0.02) & 3.79 & ($-$0.05)\tabularnewline
$\ce{H2O}$ (2$R_{e}$)  & 3.04  & 3.10  & (2.12) & 4.71 & (54.98)\tabularnewline
$\ce{OH-}$  & 1.28  & 1.28  & (0.00) & 1.28 & (0.00)\tabularnewline
$\ce{OH-}$ (2$R_{e}$)  & 23.16  & 23.16  & (0.00) & 23.16 & (0.00) \tabularnewline
$\ce{NH3}$  & 9.02  & 9.02  & (0.02) & 9.01 & ($-$0.08)\tabularnewline
$\ce{NH3}$ (2$R_{e}$)  & 13.01  & 14.27  & (9.66) & 0.01 & ($-$99.91) \tabularnewline
\end{tabular}\caption{$C_{6}^{AA}$ coefficients (in a.u.) for selected molecules (STO-3G basis) computed with FCI, UCCSD, or CCSD. Relative deviations (in percent) are given in parentheses.}
\label{tab:c6-coeffs} 
\end{table}

\section{Conclusion}\label{sec:conclusion}
In this work, we have formulated matrix-free (Davidson) methods for obtaining excitation energies and linear response properties with the q-sc-LR method.
Following a ground-state VQE calculation, our approach evaluates Hessian-vector products with arbitrary trial vectors with a finite difference scheme using gradients of the excitation operators. 
In a noise-free setting, we show that the Davidson method reproduces excitation energies of the full diagonalization approach numerically exactly but with a significantly reduced cost, mainly from needing to evaluate far fewer Hessian-vector multiplications.
We analyze some of the effects of sampling noise on calculating the Hessian vector products and find that statistical noise necessitates surprisingly large steps to handle the statistical noise. We explore the use of higher-order finite difference schemes and demonstrate that the increased stability at high step sizes allows for much larger step sizes, alleviating the sensitivity to statistical noise.
We compute linear response properties in a noise-free setting with the Davidson method. We show that static UCCSD polarizabilities match FCI reference results well. We also find that the UCCSD method outperforms classical projected CCSD for strongly correlated, stretched systems such as \ce{H2O} and \ce{NH3}.
Using damped response theory, we probe the nitrogen K-edge X-ray absorption of ammonia and find that the UCCSD spectra match the FCI reference result well with a shift of at most 0.6 eV.

\begin{acknowledgement}
We acknowledge the financial support of the Novo Nordisk Foundation for the focused research project \textit{Hybrid Quantum Chemistry on Hybrid Quantum Computers} (HQC)$^2$, grant number NNFSA220080996.
\end{acknowledgement}

\begin{suppinfo}
 Geometries of \ce{H2}, \ce{LiH}, \ce{H2O}, \ce{OH-}, and \ce{NH3} in \texttt{xyz} format. 
\end{suppinfo}

\bibliography{main}
\end{document}


\section{Geometries}
\subsection{\ce{H2}}
\begin{lstlisting}
2

H  0.0  0.0  0.0
H  0.0  0.0  0.74
\end{lstlisting}

\subsection{\ce{LiH}}
\begin{lstlisting}
2

Li  0.0  0.0  0.0
H  0.0  0.0  1.671707274
\end{lstlisting}

\subsection{Stretched \ce{LiH}}
\begin{lstlisting}
2

Li  0.0  0.0  0.0
H  0.0  0.0  3.343414548
\end{lstlisting}

\subsection{\ce{H2O}}
\begin{lstlisting}
3

O  0.0  0.0  0.1035174918
H  0.0  0.7955612117  -0.4640237459
H  0.0  -0.7955612117  -0.4640237459
\end{lstlisting}

\subsection{Stretched \ce{H2O}}
\begin{lstlisting}
3

O  0.0  0.0  0.2070349836
H  0.0  1.5911224234  -0.9280474918
H  0.0  -1.5911224234  -0.9280474918
\end{lstlisting}

\subsection{\ce{OH-}}
\begin{lstlisting}
2

O  0.0  0.0  0.0
H  0.0  0.0  1.0156985934
\end{lstlisting}

\subsection{Stretched \ce{OH-}}
\begin{lstlisting}
2

O  0.0  0.0  0.0
H  0.0  0.0  2.0313971868
\end{lstlisting}

\subsection{\ce{NH3}}
\begin{lstlisting}
4

N  0.0  0.0  -0.06972
H  0.0  -0.93223  0.32291
H  0.80734  0.46611  0.32291
H  -0.80734  0.46611  0.32291
\end{lstlisting}

\subsection{Stretched \ce{NH3}}
\begin{lstlisting}
4

N  0.0  0.0  -0.13944
H  0.0  -1.86446  0.64582
H  1.61468  0.93222  0.64582
H  -1.61468  0.93222  0.64582
\end{lstlisting}